\documentclass[%
aps,
amsmath,amssymb,
  reprint,%
]{revtex4-1}

\usepackage{subfigure}
\usepackage{hyperref}
\usepackage{float}
\usepackage{setspace}
\usepackage{afterpage}
\usepackage{amsmath}
\usepackage{placeins}
\usepackage{graphicx}
\usepackage{bm}
\usepackage{multirow}
\usepackage[varg]{txfonts}

\usepackage{caption}
\usepackage{helvet}       

\usepackage{mciteplus}

\newcommand{\degC}{\ensuremath{^{\circ}\mathrm{C}}}

\newcommand{\rmd}{\ensuremath{\mathrm{d}}}
\newcommand{\Figref}[1]{Figure~\ref{#1}}
\newcommand{\Eqref}[1]{Equation~\eqref{#1}}
\newcommand{\OrderOf}[1]{\ensuremath{{\mathcal O}\left(#1\right)}}
\newcommand{\tenpow}[1]{\ensuremath{\times10^{#1}}}

\newcommand{\Udp}{\ensuremath{U_{\mathrm{p}}}}
\newcommand{\Uc}{\ensuremath{U_{\mathrm{p}}^{\mathrm{c}}}}
\newcommand{\Ue}{\ensuremath{U_{\mathrm{p}}^{\mathrm{e}}}}
\newcommand{\Na}{\ensuremath{N_{\mathrm{A}}}}

\newcommand{\suppl}{(see Supplementary Information for details)}

\usepackage[colorinlistoftodos]{todonotes}
\newcounter{pstctr}

\begin{document}

\author{Rakhi Dhuriya} 
\author{Varun Dalia}
\author{P. Sunthar}
\email{p.sunthar@iitb.ac.in} \affiliation {Department of Chemical
  Engineering, Indian Institute of Technology Bombay (IITB), Powai,
  Mumbai, 400076, India}

\title{Diffusiophoretic Enhancement of Mass Transfer by Nanofluids}

\begin{abstract}
  Observations of an enhanced mass transfer in nanofluids have led to
  several propositions for the underlying cause, but none of them have
  been clearly established. Here, we reproduce the enhancement
  phenomenon within a glass capillary containing fluorescein di-sodium
  dye solution on one side and alumina nanoparticle suspension on the
  other, avoiding convective interferences present earlier.  The
  enhancement is explained by the counter-convective motion of the dye
  solution in response to the diffusiophoretic motion of the
  nanoparticles towards a higher concentration of the dye. The
  velocity of the dye front agrees with the theoretical estimate
  obtained from the diffusiophoretic velocity of alumina nanoparticles
  in a gradient of fluorescein di-sodium electrolyte solution.  With a
  suitably chosen nanofluid, it should now be possible to effect the
  enhancement (or suppression) of mass transfer of a given solute.
\end{abstract}


\maketitle
\textbf{ToC Graphic}
\begin{center}
  \includegraphics[width=0.8\linewidth]{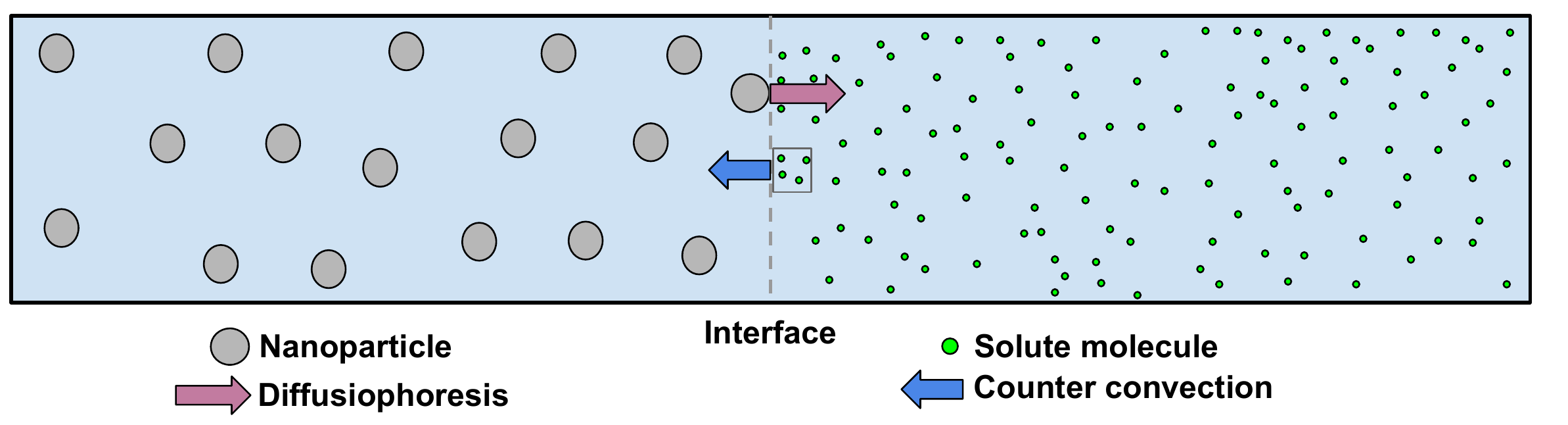}
\end{center}


\section{Introduction}

The presence of a small fraction of nanoparticles as a suspension in a
fluid (also referred to as nanofluids) seems to immensely influence
various phenomenon like diffusion \cite{Krishna2006, Fang2009,
  Veilleux2010}, absorption \cite{Komati2010, Pang2012,
  Olle2006,Esmaeili-Faraj2016, Moraveji2013,Rahmatmand2016},
extraction \cite{MirzazadehGhanadi2014, Saien2012,
  Ashrafmansouri2016}, in radiation\cite{Waheed2015}, electric
conductivity \cite{Liu2016}, evaporation \cite{Javed2013} and reaction
kinetics of chemical processes.  Mass transfer studies in
dye-diffusion by \citealt{Krishna2006} and others \cite{Veilleux2010,
  Fang2009} have demonstrated, through dramatic visual effects, that
the presence of nanoparticles increases the effective diffusivity up
to 14 folds. Yet, when carried out in other system configurations,
there are some cases where no enhancement
\cite{Ozturk2010,Subbarao2011} and even a lowering of the diffusivity
\cite{Turanov2009, Gerardi2009, Feng2012} are observed.  Various
possibilities have been advanced in an attempt to explain these
intriguing observations: Brownian motion, micro-convection
\cite{Fang2009,Krishna2006}, interfacial
complexation\cite{Ozturk2010}, dispersion model \cite{Veilleux2011},
etc., but none of them have been backed with conclusive evidence.
Here, we explore our hypothesis that the enhancement is because of a
diffusiophoretic motion \cite{Prieve1984} of the nanoparticles, a well
established concept in colloidal physics, resulting in a counter
convective motion of the dye (or other solute) molecules, leading to
an increase in the observed diffusivity of the solute.

\begin{figure*}[tbp]
  \includegraphics[width=0.8\linewidth]{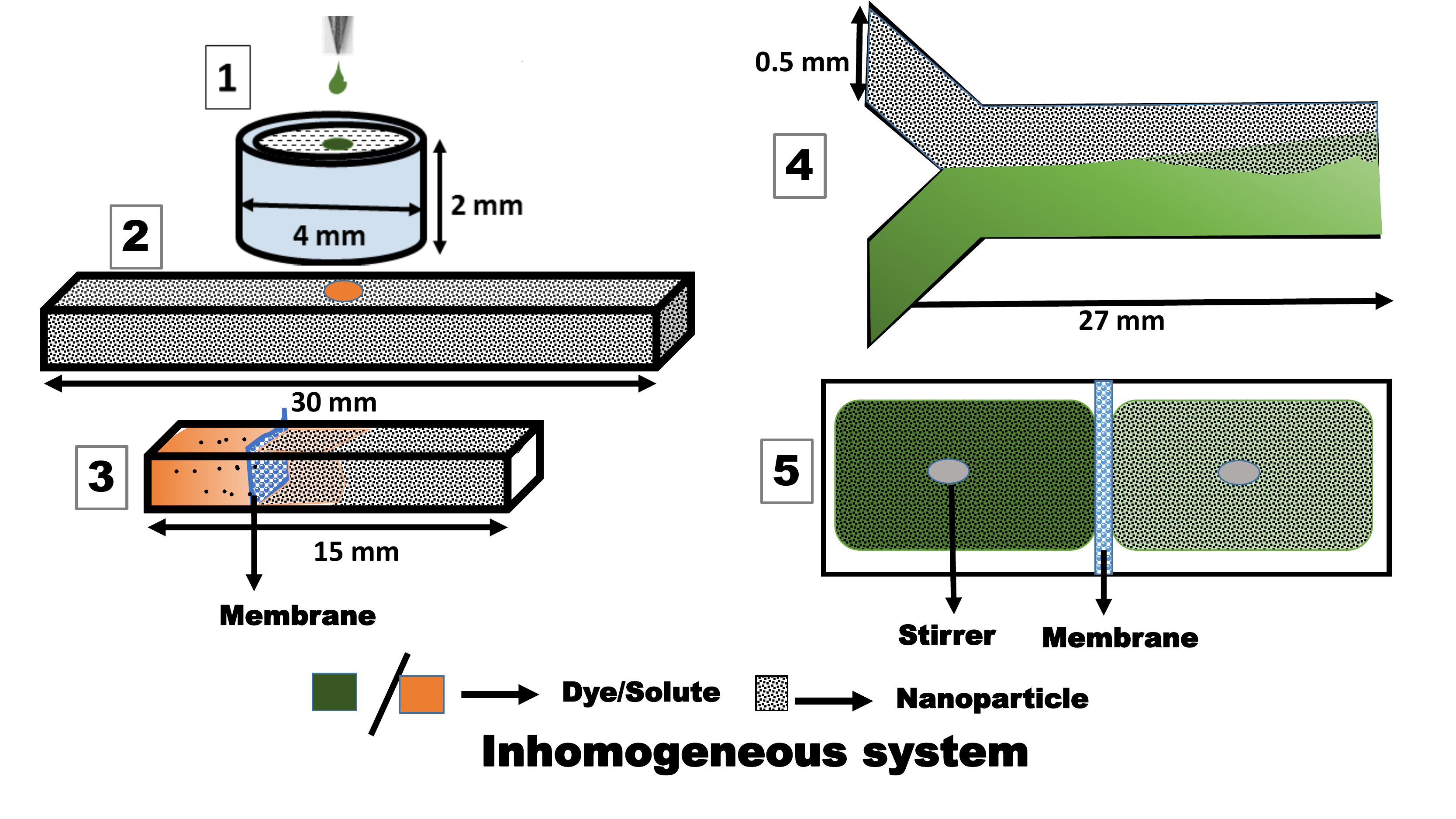}
  \caption{Schematic of a few inhomogeneous systems where the
    effect of nanoparticles on the diffusivity of a solute were
    studied.
    1: Dye drop set up of \citealt{Krishna2006}, and of 2: \citealt{Fang2009}.
    3: Membrane barrier diffusion of \citealt{Veilleux2010}. 
    4: Microfluidic setup of \citealt{Ozturk2010}, 5: Diffusion cell
    of \citealt{Feng2012}. In the last two cases, no enhancement of
    the solute diffusivity was observed.}
\label{fig:st2}
\end{figure*}

We first elaborate the contrasting observations on the effect of
nanoparticles.  For a homogeneously mixed system, where the
nanoparticles and the solute are uniformly dispersed, and the
diffusivity is measured using Fluorescence Correlation Spectroscopy
(FCS)\cite{Subbarao2011} or Nuclear magnetic resonance
(NMR)\cite{Turanov2009, Gerardi2009}, no enhancement or a small
lowering of the diffusivity of the solute has been reported. A
significant enhancement is observed only in an inhomogeneous system
(various configurations are shown in \Figref{fig:st2})
\cite{Krishna2006, Fang2009, Veilleux2010}. However, even here there
are instances where there is little enhancement \cite{Ozturk2010} or
even a decrease in the apparent diffusivity of the
solute\cite{Feng2012}.

Given the ubiquitous presence of a gradient in the dye or other solute
concentrations, we propose that the anomalous increase in diffusivity
is due to diffusiophoresis. Diffusiophoresis is the phoretic motion
of rigid colloidal particles in the presence of a gradient in the
concentration of a solute that interacts with the surface of the
colloidal particle. Diffusiophoresis is a well-known phenomenon since the
last century, and \citealt{Derjaguin1961} provided a simple expression
for the particle velocity. The most comprehensive derivation of the
equations of motion of the particles for Brownian and non-Brownian
suspensions has been recently published \cite{Brady2011} for
non-electrolyte suspensions.   Depending on the nature of the interaction,
attractive or repulsive, the particle experiences a positive or
negative velocity, respectively, along the direction of the positive
concentration gradient. The micro-mechanical origin of this velocity
is due to a solute induced gradient in the normal
stresses in the fluid surrounding the particle, that is balanced by
the viscous forces at the surface of the solid \cite{Anderson1982,
  Prieve2008}. Our postulate is that the drift motion
of nanoparticles in a concentration gradient induces a
counter-convective motion of the solvent containing the dye molecules
leading to an increase in their spread, which is often reported as an
increase in the diffusivity.

In this work we construct a system identical in the fundamental
phenomenology to the system used in the dye-drop experiment
\cite{Krishna2006}, but without the bulk convective effects resulting
from the inertial motion as the drop is dispensed in the liquid.  To
achieve this, we recognize that when a drop of dye is placed in a
suspension of nanoparticles, an interface between two nearly
``stationary'' fluids is created---one side being the solution of dye
in water and the other side is the suspension of nanoparticles.  By
using a controlled system having a similar configuration, we avoid
the irregular motion observed in the earlier work \cite{Krishna2006},
measure the initial velocity of the dye front and compare it with the
estimate obtained using the diffusiophoretic velocity of the
nanoparticles.

\section{Methodology}

\begin{figure*}[tbp]
\subfigure[]{\label{fig:capfill}\includegraphics[width=0.8\linewidth]{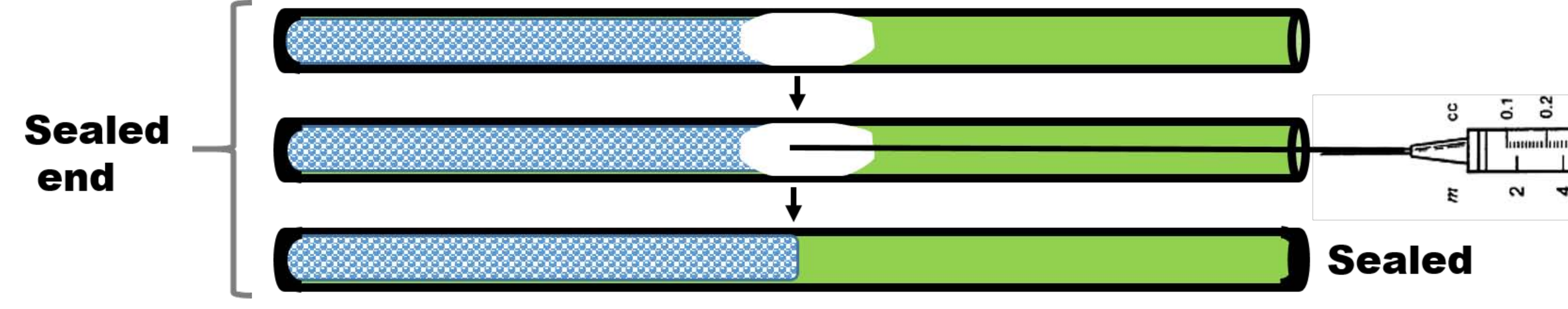}}
\subfigure[]{\label{fig:concprofile}\includegraphics[width=0.8\linewidth]{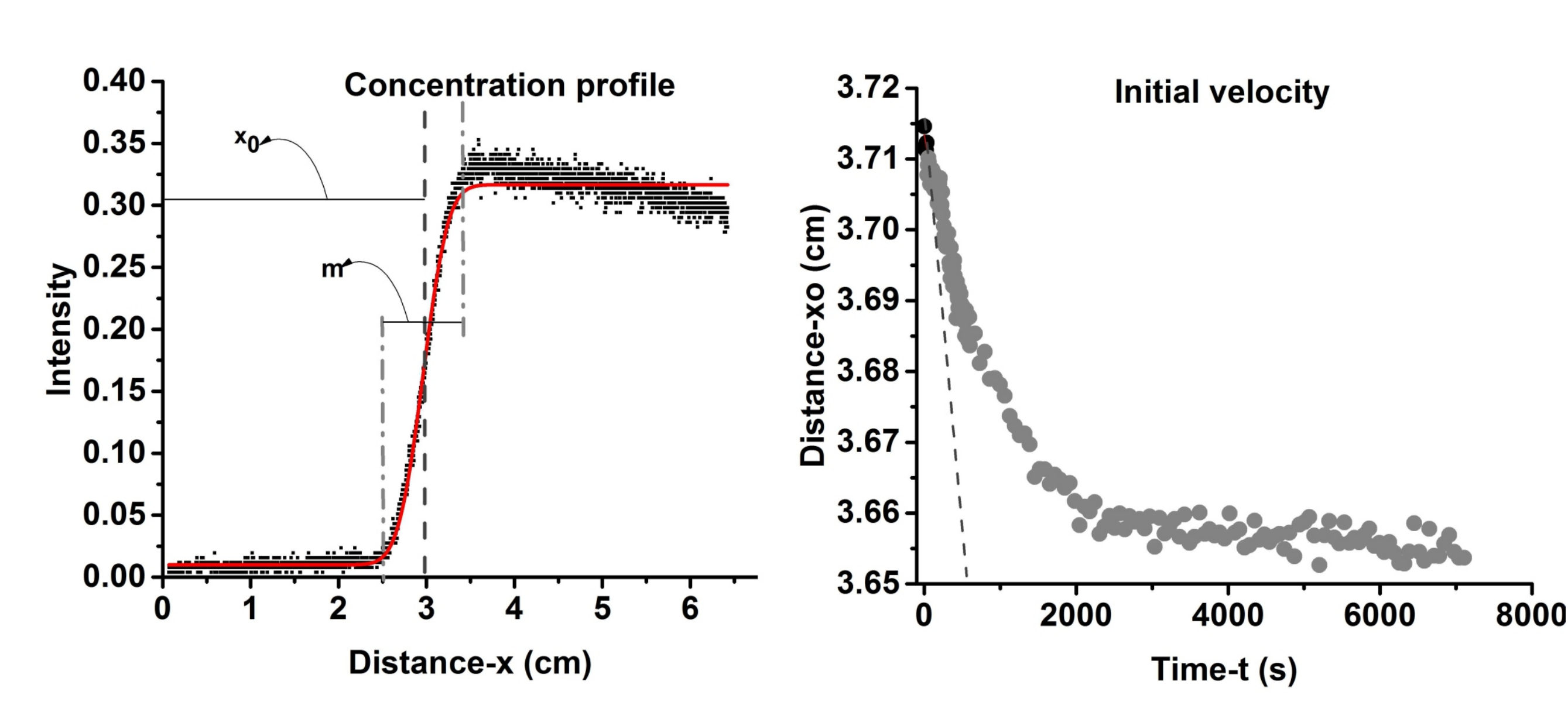}}
\subfigure[]{\label{fig:waternano}\includegraphics[width=0.8\linewidth]{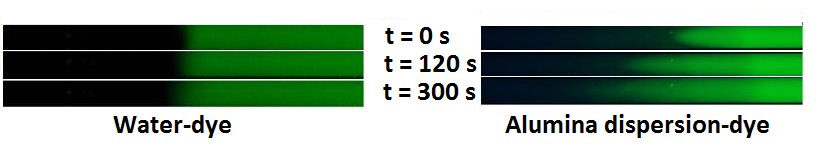}}

	%
	
\caption{(a) Schematic of steps for forming a stationary interface
  inside a capillary  (b)
  Extraction of the initial velocity from intensity profiles at
  various times  (c) Images of dye front captured at various time
  instances (in seconds) placed against water and nanofluid.} 
\label{fig:schcases}
\end{figure*}

A non-flow (one end sealed) glass capillary (I.D. 1~mm, length 6~cm)
is first filled with a dispersion of nanoparticles (alumina, 13~nm
diameter, from Sigma Aldrich) in water, as shown in the schematic in
\Figref{fig:capfill}.  This is followed by injecting a dye solution
(fluorescein di-sodium with a molecular weight of 376.27 g/mol, from
Himedia Laboratories) leaving an air gap in between. The air bubble is
then gently sucked out using a blunt needled (25 gauge) syringe,
causing the outer fluid to close upon forming a stationary nearly
straight interface (the sealed end prevents a bulk motion of the
nanofluid)\cite{VarunD2012}.  The open end is finally sealed using
paraffin wax or film.
 
The nanoparticle suspension is charge stabilized with ethane-diol
(0.5\% v/v) and NaOH ( \% v/v) at a pH of 9, which shows a greater
extent of stabilization with zeta value 39.6$\pm$1.2~mV, than the
other surfactant based systems (Zeta of alumina in water is
3.9$\pm$0.4~mV).  Earlier studies \cite{Krishna2006} have mostly used
Tween-80, but we find that the stabilization with Tween-80 is not
adequate for the horizontal capillary system: the nanoparticles settle
down during the period of the experiment, leading to secondary
convective flows which can also influence the spread of the
dye. Ethane-diol also reduces interfacial complexation with flourescein
\cite{Ozturk2010}, with the observed visual absence of intensified
fluorescent band. An equal percentage of ethane-diol--NaOH is also
present in the aqueous (dye) phase, to prevent any concentration
gradient induced effects.

This entire capillary set-up is placed inside a temperature regulated
dark box with a UV light source. Digital images are captured using
Nikon D90 DSLR (50-70mm lens, f/3.3-f/4.5), at programmed regular
intervals using a camera control software (DIYPhotobits).  The images
are processed in Matlab-R2015a image processing toolbox, where the
intensity of the green channel is extracted. The concentration of the
dye has been chosen to be below 2~mM (the critical concentration above
which dimerization of the dye leads to quenching) so that the
intensity of the fluorescently emitted light is linearly proportional
to the concentration of the dye\cite{Arbeloa1981}.

The concentration profile of the dye follows a nearly sigmoid pattern
shown in \Figref{fig:concprofile}, which can be fit to an equation
\begin{equation} 
  C = a + b \, \mathrm{erf} \left(\frac{x-x_0}{m}\right).
\label{eq:cfit}
\end{equation}%
Here, $x_{0}$ denotes the center of the interfacial region, and $m$ is
a characteristic width of the interface, $a$ and $b$ are other
constants. The dye diffusivity ($\mathcal{D}$) can be derived as
\begin{equation}
\mathcal{D}=\frac{1}{4}\,\frac{\rmd  m^2}{\rmd t}\Bigg\arrowvert_{t =
  0}.
\label{eq:diffusivity}
\end{equation}
In the absence of nanoparticles, the dye front diffuses through the
aqueous phase with ethane-diol--NaOH (this is similar to the
experiment where a dye drop is added to pure water in the earlier work
\cite{Krishna2006}). The diffusivity calculated by the above method is
found to be 600~$\pm$57~$\mu$m$^2$/s.  This is within 6-15\% of the
values reported earlier \cite{Culbertson2002,Saltzman1994,Tomaso2011},
thus validating our method of analyzing the propagation of the dye
front. The presence of ethane-diol also has negligible influence
on the diffusivity of flourescein.

\section{Results and discussion}

\begin{figure*}[tbp]
  \subfigure[]{\label{fig:npeff}\includegraphics[width=0.6\linewidth]{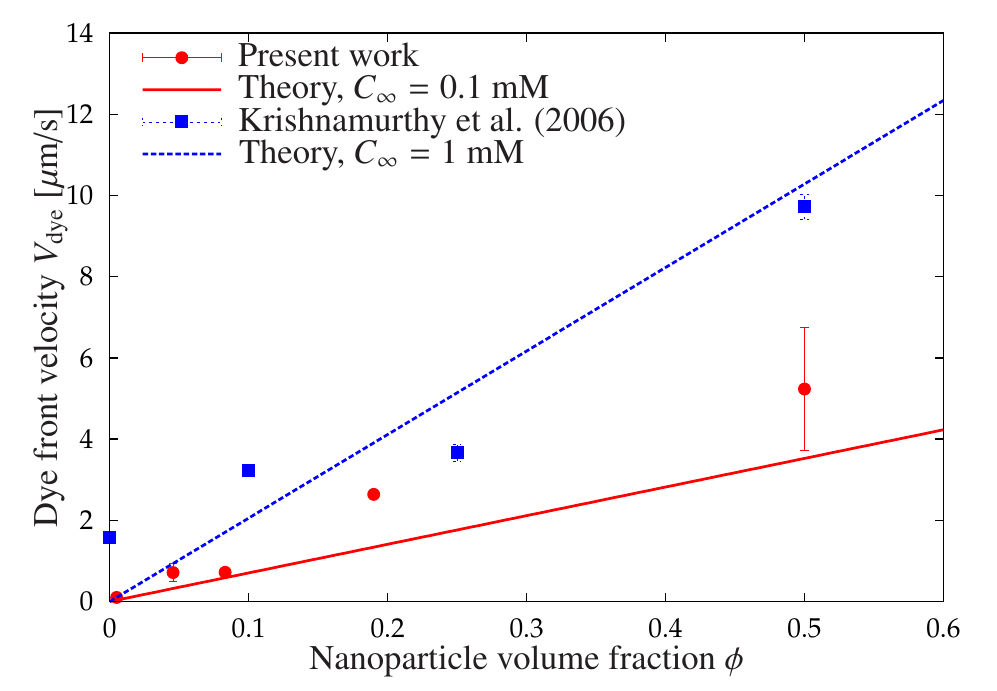}}
  \subfigure[]{\label{fig:dyeff}\includegraphics[width=0.6\linewidth]{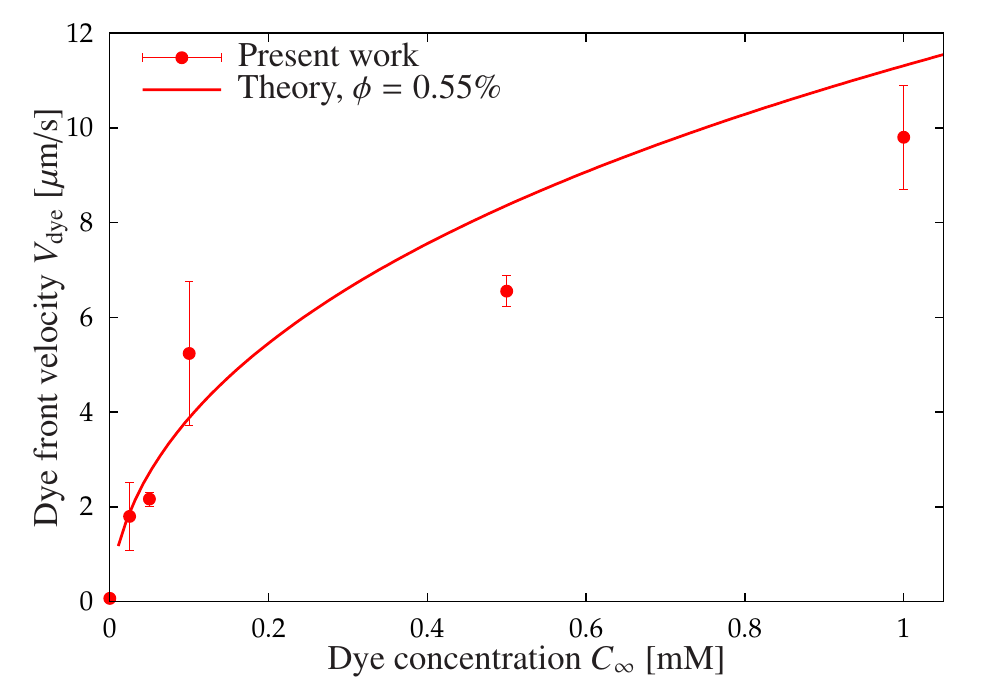}}
  \caption{Dependence of the velocity of the dye front (a) Effect of
    initial nanoparticle concentration (with an initial dye
    concentration 0.1~mM), compared with the results reported by
    \protect\citealt{Krishna2006} (with an initial dye concentration
    of 1~mM) (b) Effect of initial dye concentration at a nanoparticle
    concentration of 0.55\% (v/v). In both the plots the lines are
    the theoretical velocity evaluated from
    \protect\Eqref{eq:phidep}.}
  \label{fig:dyevel}
\end{figure*}

We now present the main results of the motion of the dye front into a
suspension of nanoparticles. When the dye front comes in contact with
the nanofluid front, the motion of the dye is relatively dramatic
(similar to the what was reported in the dye-drop experiment), as
shown in \Figref{fig:waternano} compared to the nearly still front in
the absence of alumina particles.  The time evolution of the
concentration profile is also different, that we find it better
described by a linear translation the dye front.  The front velocity
(${\rmd x_0}/{\rmd t}$) is maximum at the beginning and slowly tapers
off to zero with time, as can be inferred from
\Figref{fig:concprofile}. For the purpose of this study we limit the
discussion to the initial velocity of the dye front
\begin{equation}
V_{\mathrm{dye}}=\frac{\rmd x_0}{\rmd t}\Bigg\arrowvert_{t = 0}.
\label{eq:velocity}
\end{equation}
The effect of alumina nanoparticle concentration on $V_{\mathrm{dye}}$
for a fixed dye concentration (0.1~mM) is shown in \Figref{fig:npeff}.
The velocities are comparable to the velocities obtained in the
dye-drop experiment \cite{Krishna2006}, implying the phenomenon
occurring in the horizontal capillary system is similar to the former.
The effect of dye concentration on $V_{\mathrm{dye}}$, shown in
\Figref{fig:dyeff}, has not been reported in the earlier
experiments. This is also crucial to infer the nature of the driving.

Diffusiophoresis can explain the enhanced motion observed in
\Figref{fig:dyevel}. Fluorescein (which is the solute here) is known
to interact with alumina nanoparticles forming a surface complex
\cite{Ozturk2010} (in essence an attractive interaction). When the
nanofluid phase comes in contact with the dye phase, the nanoparticles
at the interface are effectively placed in a steep gradient, with the
dye only on one side of the nanoparticles at the interface.  This
causes a motion of the nanoparticles towards the dye phase by
diffusiophoresis (towards a higher concentration of the dye due to the
attractive interaction). Since the ends of the capillary are sealed,
incompressibility of the fluid implies that an equal volume of dye
bearing fluid will move towards the nanofluid phase, which is visually
observed as an enhanced motion, more than the normal dye diffusion.

This behavior can be quantified more precisely using the
diffusiophoretic velocity experienced by a nanoparticle placed in a
gradient of an electrolyte. Whereas the expressions derived in the
references cited earlier \cite{Anderson1982,Brady2011} deal with a
case when the solute is a non-electrolyte, the presence of ionized
solute includes an additional contribution from an
electrophoretic effect: motion in a potential generated due to
differential diffusivities of the solute ions\cite{Prieve2008}.  The
diffusiophoretic velocity of a rigid particle in an arbitrary
zeta-potential \cite{Prieve1984} and an arbitrary Debye length of the
counterion cloud has been derived \cite{Prieve1987}. For brevity of the
discussion we quote only the necessary expressions here, with the
detailed expressions given elsewhere \suppl.  The particle velocity
arises due to two contributions
\begin{equation}
  \label{eq:Udp}
  \Udp = \Uc + \Ue,
\end{equation}
where, \Uc\ is the chemophoretic component (similar to that derived
for non-electrolyte solutes \cite{Anderson1982,Brady2011}), and \Ue\
is the electrophoretic component. The expressions have been derived in
for weakly non-uniform solute gradient
($a \, |\nabla \ln C_{\infty}| \ll 1$) in the asymptotic limit of the
Debye length being small compared to the particle radius
\begin{equation}
 \lambda \equiv \frac{\kappa^{-1}}{a} \ll 1.
\end{equation}
In the leading order the velocities are given by
\begin{align}
  \label{eq:Uc}
  \Uc &=  \frac{\varepsilon}{2\pi\, \eta}\left(\frac{kT}{Z\,e}
  \right)^2 \left[- \ln(1-\mathrm{\gamma}^2) \right]
        \mathrm{\nabla}\ln C_{\infty}
        + \OrderOf{\lambda}, \\
  \label{eq:Ue}
  \Ue &=  \frac{\varepsilon}{4\pi\, \eta}\left(\frac{kT}{Ze}
  \right)^2\beta \, \bar{\zeta} \, \mathrm{\nabla}\ln C_{\infty}
        + \OrderOf{\lambda}
\intertext{with,} 
  \varepsilon &\equiv 4 \, \pi \, \varepsilon_{0} \,
                \varepsilon_{\mathrm{r}} \\
  \gamma &\equiv \tanh \frac{\bar\zeta}{4} \\
  \bar{\zeta} & \equiv \frac{Z \, e \,\zeta}{kT} \qquad
                \text{(normalized zeta potential)}\\
  \beta &\equiv\frac{D_+ -D_-}{Z_+ \, D_+ - Z_{-} \, D_{-}}
                   \qquad \text{(normalized difference in the
                   diffusivities)}\\
  \kappa&\equiv\left(\frac{4 \pi e^2\, Z^{2} \, \Na \, C_{\infty}
          }{\varepsilon \, kT} \right)^{1/2}\qquad 
          \text{(inverse Debye length)}
\end{align}
where, $\varepsilon_{0}$ is the permittivity of free-space,
$\varepsilon_{\mathrm{r}}$ is the relative permittivity of the medium,
$\eta$ is the viscosity of the medium, $k$ is the Boltzmann constant,
$T$ is the temperature, $Z$ is the valency of the ions of the assumed
symmetric electrolyte---i.e., equal for positive and negative ions
$Z = Z_{+} = -Z_{-}$, $e$ is the elementary positive charge, $\zeta$
is the zeta-potential at the particle surface, $D_{+}$ and $D_{-}$ are
the diffusion coefficients of the positive and negative ions
respectively, $\Na$ is the Avogadro number, and $C_{\infty}$ is the
concentration (in mM) of the solute far from the particle surface. $n$
is the number of species. Higher order velocity corrections in
$\lambda$ have been derived, and require a numerical evaluation of
certain functions \cite{Prieve1984} \suppl.

A brief interpretation of the driving forces and their resultant
effect on the particle velocity follows. The term in the square
brackets in \Eqref{eq:Uc} is always positive, implying the
chemophoretic velocity \Uc\ is always in the direction of the
concentration gradient, irrespective of the sign of the zeta
potential.  An equivalent physical interpretation would be: since the
diffuse layer is oppositely charged to the surface, there is always an
``attractive'' interaction between the solute molecules and the
surface, leading to the particle moving in a direction towards a higher
concentration (as in the case of non-electrolyte diffusiophoresis).

The electrophoretic velocity component \Ue\ is due to a net electric
field generated because of different diffusivities of the two
oppositely charged solute ions \cite{Prieve2008}. Consider a case of
$D_{+} > D_{-}$ or $\beta>0$. This sets up an electric field in the
direction of increasing solute concentration ($C_{\infty}$), because
the positive ions diffuse faster than the negative ions. A positively
charged particle ($\zeta >0$) will, therefore, move towards regions of
higher solute concentrations, and a negatively charged particle
towards lower concentrations.  For a system with the product
$(\beta \, \zeta)>0$, the particle will always move towards a higher
concentration of the solute. In the opposite case, the direction of
the net diffusiophoretic velocity depends on competition between a
positive chemophoretic velocity and a negative electrophoretic
velocity.

We can now estimate the diffusiophoretic velocity of a nanoparticle in
a gradient of fluorescein-di-sodium.  The sodium salt of fluorescein
has a solubility of 500~g/L (1.33~M) at 20\degC, implying that in the
present system (with concentrations in mM) the salt is completely
ionized. For the sodium ion
$D_{+} = 13.3\tenpow{-10}$~m$^{2}$/s\cite{Cussler2009}, and for
fluorescein $D_{-} = 3 \tenpow{-10}$~m$^{2}$/s, implying $\beta >0$.
The zeta potential of the alumina particles is positive and is a
function of the concentration of the dye \suppl. The valency of
fluorescein is taken to be $Z_{-} = 2$ at a pH $\approx 9$
\cite{Roberto2010}, and the valency of sodium is $Z_{+} = 1$. In the
expressions leading to \Eqref{eq:Udp}, which have been derived for a
symmetric electrolyte, we take $Z=2$ (magnitude of the highest valency
of the counterions). This is also the correct limit for a highly
charged particle, the potential around which is determined by the
counterion with the largest valency \cite{obrien83}.  Using other
standard values at a temperature $T = 298$~K and the higher order
corrections in $\lambda$ \suppl, the theoretical diffusiophoretic
velocity can be calculated using \Eqref{eq:Udp}.  Since it is the
initial velocity that is experimentally measured, when one side of a
nanoparticle at the interface has the dye, and the other is devoid of
it, we approximate the concentration gradient to
$\nabla\ln C_{\infty} \approx 1/(2 \,a) $ where $a$ is the radius of
the particle.  Though this is contrary to the assumption of a weakly
non-uniform solute gradient, this is the best estimate we can get
\cite{Brady2011}. The particle velocity is converted to a velocity of
the dye front $V_{\mathrm{dye}}$ from the condition that there is no
net flow in the capillary across any cross section (since the ends are
sealed). This gives
\begin{align}
  \label{eq:phidep}
  V_{\mathrm{dye}} = - \frac{\phi}{1-\phi} \, \Udp \approx -\phi \,
  \Udp,
\end{align}
where $\phi$ is the volume fraction of the nanoparticle suspension,
and the approxmiation is valid for a dilute suspension $\phi \ll 1$.
The theoretical velocity of the dye front from \Eqref{eq:phidep} is
also plotted in \Figref{fig:dyevel}. This independent theoretical
estimate (without any adjustable parameters) captures the dependence
on the nanoparticle concentration (\Figref{fig:npeff}) as well as the
dye concentration (\Figref{fig:dyeff}).  We also find good agreement
with the velocity calculated from the radial spread of the dye
reported in the dye-drop experiment \cite{Krishna2006}
(\Figref{fig:npeff}, in spite of the asymptotic nature of the solution
\cite{Prieve1984} and the estimation approximations we have made.

The dependence of the velocity on the concentration of the nanofluid
suspension (in \Figref{fig:npeff}) arises (in the dilute limit) merely
because of \Eqref{eq:phidep}.  However, the dependence on the
concentration of the solute enters in a non-trivial manner, which is
very different from that of a non-electrolyte solute. For
non-electrolyte solutes the diffusiophoretic velocity is directly
proportional to the concentration of the solute through the relation
\cite{Anderson1984,Brady2011}
\begin{equation}
  \Udp \sim \delta^2 \frac{kT}{\eta} \, \nabla C_{\infty},
  \label{eq:udpnon}
\end{equation} 
where, $\delta$ is a characteristic length of the solute-particle
interaction potential. In contrast for electrolyte gradients, in
\Eqref{eq:Udp}, the concentration gradient dependence is weak
($ \sim \nabla \ln C_{\infty}$).  Nevertheless, the bulk concentration
$C_{\infty}$ influences other parameters such as the zeta potential
$\bar\zeta$ and the Debye length $\kappa^{-1}$, leading to a
significant effect on the particle velocity $\Udp$ seen in
\Figref{fig:dyeff}.

The irregular spread of the dye in the dye-drop experiment
\citep{Krishna2006} needs further investigation.  It is possible that
this is due to an instability in the flow resulting from the
counter-convective motion. Within the capillary, we did not observe any
irregular motion, indicating a possible suppression of the growth of
the instability within the confinement.

\section{Conclusion}

To conclude, the horizontal capillary system and the method to create
a stationary interface, between the nanoparticle suspension on one
side and dye on the other side, reproduces the behavior of velocities
observed in the dye-drop experiments \cite{Krishna2006}.  The initial
velocities measured in these systems can be quantitatively explained
by diffusiophoresis of large particles leading to a counter-convective
motion of the dye bearing solvent. Diffusiophoresis also corroborates
with other experiments where the ``enhanced mass transfer'' occurs
only when the system is initially out of equilibrium (gradient in the
solute concentration), and not when it is homogeneous.  The
explanation of the absence of enhancement in a few cases
\cite{Feng2012} needs further work: studies with other systems
(solutes and nanoparticles), given that diffusiophoretic velocities
have been measured in similar diffusion/diaphragm cells
\cite{Lechshae1984}.  The important takeaway is that there is
enhancement of mass transfer ``by'' nanofluids rather than ``in''
nanofluids, and this can be tuned. By choosing or altering the
solute-nanoparticle interactions, it should now be possible to enhance
micro-mixing or supress it. In micro-enviroments requiring fast
mixing, such as in nanofabrication, lab-on-chip, drug-delivery,
chemotaxis, etc., the effects can be substantial.
	
\section*{Acknowledgements}
  We acknowledge the use of facilities in SAIF and CRNTS, IIT Bombay
  used in this work, and a funding from SERB-DST, Govt. of India, for
  developing in-house facilities in our laboratory.

\bibliography{nanodiff}

\end{document}